%Paper: hep-ph/9307342
%From: SHER@WMHEG.PHYSICS.WM.EDU
%Date: Tue, 27 Jul 1993 8:44:33 -0400 (EDT)

%  THE FIRST 850 LINES CONSIST OF THE PHYZZX MACRO PACKAGE.
%  THIS SHOULD RUN UNDER PLAIN TEX
%
%
\vsize=8.75truein
%%%%%%%%%%%%%%%%%%%%%%%%%%%%%%%%%%%%%%%%%%%%%%%%%%%%%%%%%%%%%%%%%%%%%%%%
% % % % % % % % % % % % % % % % % % % % % % % % % % % % % % % % % % % %
%%%   This is PHYZZX macro package.   % % % % % % % % % % % % % % % % %
%% % % % % % % % % % % % % % % % % % % % % % % % % % % % % % % % % % % %
%%%  This version of PHYZZX should be used with Version 1.0 of TEX  % %
%% % % % % % % % % % % % % % % % % % % % % % % % % % % % % % % % % % % %
%%%%%%%%%%%%%%%%%%%%%%%%%%%%%%%%%%%%%%%%%%%%%%%%%%%%%%%%%%%%%%%%%%%%%%%%
%%%%%%%  Created by Vadim Kaplunovsky in June 1984.   %%%%%%%%%%%%%%%%%%
% % % % % % % % % % % % % % % % % % % % % % % % % % % % % % % % % % % %
%%%%%%%%%%%%  Latest update/debug: September 24, 1984.    %%%%%%%%%%%%%%
%%%%%%%%%%%%%%%%%%%%%%%%%%%%%%%%%%%%%%%%%%%%%%%%%%%%%%%%%%%%%%%%%%%%%%%%
%
\catcode`@=11 % This allows us to modify PLAIN macros.
%
%%%%%%%%%%%%%%%%%%%%%%%%%%%%%%%%%%%%%%%%%%%%%%%%%%%%%%%%%%%%%%%%%%%%%%%%
%
%   I begin with fonts.
%

\font\fourteenrm=cmr10 scaled\magstep2
\font\twelverm=cmr10 scaled\magstep1
\font\ninerm=cmr9            \font\sixrm=cmr6

\font\fourteenbf=cmbx10 scaled\magstep2
\font\twelvebf=cmbx10 scaled\magstep1
\font\ninebf=cmbx9            \font\sixbf=cmbx6
\font\seventeeni=cmmi10 scaled\magstep3     \skewchar\seventeeni='177
\font\fourteeni=cmmi10 scaled\magstep2      \skewchar\fourteeni='177
\font\twelvei=cmmi10 scaled\magstep1        \skewchar\twelvei='177
\font\ninei=cmmi9                           \skewchar\ninei='177
\font\sixi=cmmi6                            \skewchar\sixi='177
\font\seventeensy=cmsy10 scaled\magstep3    \skewchar\seventeensy='60
\font\fourteensy=cmsy10 scaled\magstep2     \skewchar\fourteensy='60
\font\twelvesy=cmsy10 scaled\magstep1       \skewchar\twelvesy='60
\font\ninesy=cmsy9                          \skewchar\ninesy='60
\font\sixsy=cmsy6                           \skewchar\sixsy='60

\font\fourteenex=cmex10 scaled\magstep2
\font\twelveex=cmex10 scaled\magstep1

\font\fourteensl=cmsl10 scaled\magstep2
\font\twelvesl=cmsl10 scaled\magstep1
\font\ninesl=cmsl9

\font\fourteenit=cmti10 scaled\magstep2
\font\twelveit=cmti10 scaled\magstep1
\font\twelvett=cmtt10 scaled\magstep1
\font\twelvecp=cmcsc10 scaled\magstep1
\font\tencp=cmcsc10
\newfam\cpfam
%
      % quick fix for a missing font
%
\newcount\f@ntkey            \f@ntkey=0
\def\samef@nt{\relax \ifcase\f@ntkey \rm \or\oldstyle \or\or
         \or\it \or\sl \or\bf \or\tt \or\caps \fi }
\def\fourteenpoint{\relax
    \textfont0=\fourteenrm          \scriptfont0=\tenrm
    \scriptscriptfont0=\sevenrm
     \def\rm{\fam0 \fourteenrm \f@ntkey=0 }\relax
    \textfont1=\fourteeni           \scriptfont1=\teni
    \scriptscriptfont1=\seveni
     \def\oldstyle{\fam1 \fourteeni\f@ntkey=1 }\relax
    \textfont2=\fourteensy          \scriptfont2=\tensy
    \scriptscriptfont2=\sevensy
    \textfont3=\fourteenex     \scriptfont3=\fourteenex
    \scriptscriptfont3=\fourteenex
    \def\it{\fam\itfam \fourteenit\f@ntkey=4 }\textfont\itfam=\fourteenit
    \def\sl{\fam\slfam \fourteensl\f@ntkey=5 }\textfont\slfam=\fourteensl
    \scriptfont\slfam=\tensl
    \def\bf{\fam\bffam \fourteenbf\f@ntkey=6 }\textfont\bffam=\fourteenbf
    \scriptfont\bffam=\tenbf     \scriptscriptfont\bffam=\sevenbf
    \def\tt{\fam\ttfam \twelvett \f@ntkey=7 }\textfont\ttfam=\twelvett
    \h@big=11.9\p@{} \h@Big=16.1\p@{} \h@bigg=20.3\p@{} \h@Bigg=24.5\p@{}
    \def\caps{\fam\cpfam \twelvecp \f@ntkey=8 }\textfont\cpfam=\twelvecp
    \setbox\strutbox=\hbox{\vrule height 12pt depth 5pt width\z@}
    \samef@nt}
\def\twelvepoint{\relax
    \textfont0=\twelverm          \scriptfont0=\ninerm
    \scriptscriptfont0=\sixrm
     \def\rm{\fam0 \twelverm \f@ntkey=0 }\relax
    \textfont1=\twelvei           \scriptfont1=\ninei
    \scriptscriptfont1=\sixi
     \def\oldstyle{\fam1 \twelvei\f@ntkey=1 }\relax
    \textfont2=\twelvesy          \scriptfont2=\ninesy
    \scriptscriptfont2=\sixsy
    \textfont3=\twelveex          \scriptfont3=\twelveex
    \scriptscriptfont3=\twelveex
    \def\it{\fam\itfam \twelveit \f@ntkey=4 }\textfont\itfam=\twelveit
    \def\sl{\fam\slfam \twelvesl \f@ntkey=5 }\textfont\slfam=\twelvesl
    \scriptfont\slfam=\ninesl
    \def\bf{\fam\bffam \twelvebf \f@ntkey=6 }\textfont\bffam=\twelvebf
    \scriptfont\bffam=\ninebf     \scriptscriptfont\bffam=\sixbf
    \def\tt{\fam\ttfam \twelvett \f@ntkey=7 }\textfont\ttfam=\twelvett
    \h@big=10.2\p@{}
    \h@Big=13.8\p@{}
    \h@bigg=17.4\p@{}
    \h@Bigg=21.0\p@{}
    \def\caps{\fam\cpfam \twelvecp \f@ntkey=8 }\textfont\cpfam=\twelvecp
    \setbox\strutbox=\hbox{\vrule height 10pt depth 4pt width\z@}
    \samef@nt}
\def\tenpoint{\relax
    \textfont0=\tenrm          \scriptfont0=\sevenrm
    \scriptscriptfont0=\fiverm
    \def\rm{\fam0 \tenrm \f@ntkey=0 }\relax
    \textfont1=\teni           \scriptfont1=\seveni
    \scriptscriptfont1=\fivei
    \def\oldstyle{\fam1 \teni \f@ntkey=1 }\relax
    \textfont2=\tensy          \scriptfont2=\sevensy
    \scriptscriptfont2=\fivesy
    \textfont3=\tenex          \scriptfont3=\tenex
    \scriptscriptfont3=\tenex
    \def\it{\fam\itfam \tenit \f@ntkey=4 }\textfont\itfam=\tenit
    \def\sl{\fam\slfam \tensl \f@ntkey=5 }\textfont\slfam=\tensl
    \def\bf{\fam\bffam \tenbf \f@ntkey=6 }\textfont\bffam=\tenbf
    \scriptfont\bffam=\sevenbf     \scriptscriptfont\bffam=\fivebf
    \def\tt{\fam\ttfam \tentt \f@ntkey=7 }\textfont\ttfam=\tentt
    \def\caps{\fam\cpfam \tencp \f@ntkey=8 }\textfont\cpfam=\tencp
    \setbox\strutbox=\hbox{\vrule height 8.5pt depth 3.5pt width\z@}
    \samef@nt}
%
%%%%%%%%%%%%%%%%%%%%%%%%%%%%%%%%%%%%%%%%%%%%%%%%%%%%%%%%%%%%%%%%%%%%%%%%
%
%   Next redifine \big \Big \bigg and \Bigg to work with all fonts
%
%%%%%%%%%%%%%%%%%%%%%%%%%%%%%%%%%%%%%%%%%%%%%%%%%%%%%%%%%%%%%%%%%%%%%%%%
%
\newdimen\h@big  \h@big=8.5\p@
\newdimen\h@Big  \h@Big=11.5\p@
\newdimen\h@bigg  \h@bigg=14.5\p@
\newdimen\h@Bigg  \h@Bigg=17.5\p@
\def\big#1{{\hbox{$\left#1\vbox to\h@big{}\right.\n@space$}}}
\def\Big#1{{\hbox{$\left#1\vbox to\h@Big{}\right.\n@space$}}}
\def\bigg#1{{\hbox{$\left#1\vbox to\h@bigg{}\right.\n@space$}}}
\def\Bigg#1{{\hbox{$\left#1\vbox to\h@Bigg{}\right.\n@space$}}}
%
%%%%%%%%%%%%%%%%%%%%%%%%%%%%%%%%%%%%%%%%%%%%%%%%%%%%%%%%%%%%%%%%%%%%%%%%
%
%   Next, I define basic spacing parameters.
%
\normalbaselineskip = 20pt plus 0.2pt minus 0.1pt
\normallineskip = 1.5pt plus 0.1pt minus 0.1pt
\normallineskiplimit = 1.5pt
\newskip\normaldisplayskip
\normaldisplayskip = 20pt plus 5pt minus 10pt
\newskip\normaldispshortskip
\normaldispshortskip = 6pt plus 5pt
\newskip\normalparskip
\normalparskip = 6pt plus 2pt minus 1pt
\newskip\skipregister
\skipregister = 5pt plus 2pt minus 1.5pt
\newif\ifsingl@    \newif\ifdoubl@
\newif\iftwelv@    \twelv@true
\def\singlespace{\singl@true\doubl@false\spaces@t}
\def\doublespace{\singl@false\doubl@true\spaces@t}
\def\normalspace{\singl@false\doubl@false\spaces@t}
\def\Tenpoint{\tenpoint\twelv@false\spaces@t}
\def\Twelvepoint{\twelvepoint\twelv@true\spaces@t}
\def\spaces@t{\relax%
 \iftwelv@ \ifsingl@\subspaces@t3:4;\else\subspaces@t1:1;\fi%
 \else \ifsingl@\subspaces@t3:5;\else\subspaces@t4:5;\fi \fi%
 \ifdoubl@ \multiply\baselineskip by 5%
 \divide\baselineskip by 4 \fi \unskip}
\def\subspaces@t#1:#2;{
      \baselineskip = \normalbaselineskip
      \multiply\baselineskip by #1 \divide\baselineskip by #2
      \lineskip = \normallineskip
      \multiply\lineskip by #1 \divide\lineskip by #2
      \lineskiplimit = \normallineskiplimit
      \multiply\lineskiplimit by #1 \divide\lineskiplimit by #2
      \parskip = \normalparskip
      \multiply\parskip by #1 \divide\parskip by #2
      \abovedisplayskip = \normaldisplayskip
      \multiply\abovedisplayskip by #1 \divide\abovedisplayskip by #2
      \belowdisplayskip = \abovedisplayskip
      \abovedisplayshortskip = \normaldispshortskip
      \multiply\abovedisplayshortskip by #1
        \divide\abovedisplayshortskip by #2
      \belowdisplayshortskip = \abovedisplayshortskip
      \advance\belowdisplayshortskip by \belowdisplayskip
      \divide\belowdisplayshortskip by 2
      \smallskipamount = \skipregister
      \multiply\smallskipamount by #1 \divide\smallskipamount by #2
      \medskipamount = \smallskipamount \multiply\medskipamount by 2
      \bigskipamount = \smallskipamount \multiply\bigskipamount by 4 }
\def\normalbaselines{ \baselineskip=\normalbaselineskip
   \lineskip=\normallineskip \lineskiplimit=\normallineskip
   \iftwelv@\else \multiply\baselineskip by 4 \divide\baselineskip by 5
     \multiply\lineskiplimit by 4 \divide\lineskiplimit by 5
     \multiply\lineskip by 4 \divide\lineskip by 5 \fi }
\Twelvepoint  % That's the default
\interlinepenalty=50
\interfootnotelinepenalty=5000
\predisplaypenalty=9000
\postdisplaypenalty=500
\hfuzz=1pt
\vfuzz=0.2pt
%
%%%%%%%%%%%%%%%%%%%%%%%%%%%%%%%%%%%%%%%%%%%%%%%%%%%%%%%%%%%%%%%%%%%%%%%%
%
%   Next, I define output routines, footnotes & related stuff.
%
\def\pagecontents{
   \ifvoid\topins\else\unvbox\topins\vskip\skip\topins\fi
   \dimen@ = \dp255 \unvbox255
   \ifvoid\footins\else\vskip\skip\footins\footrule\unvbox\footins\fi
   \ifr@ggedbottom \kern-\dimen@ \vfil \fi }
\def\makeheadline{\vbox to 0pt{ \skip@=\topskip
      \advance\skip@ by -12pt \advance\skip@ by -2\normalbaselineskip
      \vskip\skip@ \line{\vbox to 12pt{}\the\headline} \vss
      }\nointerlineskip}
\def\makefootline{\baselineskip = 1.5\normalbaselineskip
                 \line{\the\footline}}
\newif\iffrontpage
\newif\ifletterstyle
\newif\ifp@genum
\def\nopagenumbers{\p@genumfalse}
\def\pagenumbers{\p@genumtrue}
\pagenumbers
\newtoks\paperheadline
\newtoks\letterheadline
\newtoks\letterfrontheadline
\newtoks\lettermainheadline
\newtoks\paperfootline
\newtoks\letterfootline
\newtoks\date
\footline={\ifletterstyle\the\letterfootline\else\the\paperfootline\fi}
\paperfootline={\hss\iffrontpage\else\ifp@genum\tenrm\folio\hss\fi\fi}
\letterfootline={\hfil}
\headline={\ifletterstyle\the\letterheadline\else\the\paperheadline\fi}
\paperheadline={\hfil}
\letterheadline{\iffrontpage\the\letterfrontheadline
     \else\the\lettermainheadline\fi}
\lettermainheadline={\rm\ifp@genum page \ \folio\fi\hfil\the\date}
\def\monthname{\relax\ifcase\month 0/\or January\or February\or
   March\or April\or May\or June\or July\or August\or September\or
   October\or November\or December\else\number\month/\fi}
\date={\monthname\ \number\day, \number\year}
\countdef\pagenumber=1  \pagenumber=1
\def\advancepageno{\global\advance\pageno by 1
   \ifnum\pagenumber<0 \global\advance\pagenumber by -1
    \else\global\advance\pagenumber by 1 \fi \global\frontpagefalse }
\def\folio{\ifnum\pagenumber<0 \romannumeral-\pagenumber
           \else \number\pagenumber \fi }
\def\footrule{\dimen@=\prevdepth\nointerlineskip
   \vbox to 0pt{\vskip -0.25\baselineskip \hrule width 0.35\hsize \vss}
   \prevdepth=\dimen@ }
\newtoks\foottokens
\foottokens={\Tenpoint\singlespace}
\newdimen\footindent
\footindent=24pt
\def\vfootnote#1{\insert\footins\bgroup  \the\foottokens
   \interlinepenalty=\interfootnotelinepenalty \floatingpenalty=20000
   \splittopskip=\ht\strutbox \boxmaxdepth=\dp\strutbox
   \leftskip=\footindent \rightskip=\z@skip
   \parindent=0.5\footindent \parfillskip=0pt plus 1fil
   \spaceskip=\z@skip \xspaceskip=\z@skip
   \Textindent{$ #1 $}\footstrut\futurelet\next\fo@t}
\def\Textindent#1{\noindent\llap{#1\enspace}\ignorespaces}
\def\footnote#1{\attach{#1}\vfootnote{#1}}

\def\foot{\attach\footsymbolgen\vfootnote{\footsymbol}}
\let\footsymbol=\star
\newcount\lastf@@t           \lastf@@t=-1
\newcount\footsymbolcount    \footsymbolcount=0
\newif\ifPhysRev
\def\footsymbolgen{\relax \ifPhysRev \iffrontpage \NPsymbolgen\else
      \PRsymbolgen\fi \else \NPsymbolgen\fi
   \global\lastf@@t=\pageno \footsymbol }
\def\NPsymbolgen{\ifnum\footsymbolcount<0 \global\footsymbolcount=0\fi
   {\iffrontpage \else \advance\lastf@@t by 1 \fi
    \ifnum\lastf@@t<\pageno \global\footsymbolcount=0
     \else \global\advance\footsymbolcount by 1 \fi }
   \ifcase\footsymbolcount \fd@f\star\or \fd@f\dagger\or \fd@f\ast\or
    \fd@f\ddagger\or \fd@f\natural\or \fd@f\diamond\or \fd@f\bullet\or
    \fd@f\nabla\else \fd@f\dagger\global\footsymbolcount=0 \fi }
\def\fd@f#1{\xdef\footsymbol{#1}}
\def\PRsymbolgen{\ifnum\footsymbolcount>0 \global\footsymbolcount=0\fi
      \global\advance\footsymbolcount by -1
      \xdef\footsymbol{\sharp\number-\footsymbolcount} }
\def\space@ver#1{\let\@sf=\empty \ifmmode #1\else \ifhmode
   \edef\@sf{\spacefactor=\the\spacefactor}\unskip${}#1$\relax\fi\fi}
\def\attach#1{\space@ver{\strut^{\mkern 2mu #1} }\@sf\ }
%
%%%%%%%%%%%%%%%%%%%%%%%%%%%%%%%%%%%%%%%%%%%%%%%%%%%%%%%%%%%%%%%%%%%%%%%%
%
%   Here come chapter, section, subsection & appendix macros.
%
\newcount\chapternumber      \chapternumber=0
\newcount\sectionnumber      \sectionnumber=0
\newcount\equanumber         \equanumber=0
\let\chapterlabel=0
\newtoks\chapterstyle        \chapterstyle={\Number}
\newskip\chapterskip         \chapterskip=\bigskipamount
\newskip\sectionskip         \sectionskip=\medskipamount
\newskip\headskip            \headskip=8pt plus 3pt minus 3pt
\newdimen\chapterminspace    \chapterminspace=15pc
\newdimen\sectionminspace    \sectionminspace=10pc
\newdimen\referenceminspace  \referenceminspace=25pc
\def\chapterreset{\global\advance\chapternumber by 1
   \ifnum\the\equanumber<0 \else\global\equanumber=0\fi
   \sectionnumber=0 \makel@bel}
\def\makel@bel{\xdef\chapterlabel{%
\the\chapterstyle{\the\chapternumber}.}}
\def\sectionlabel{\number\sectionnumber \quad }
\def\alphabetic#1{\count255='140 \advance\count255 by #1\char\count255}
\def\Alphabetic#1{\count255='100 \advance\count255 by #1\char\count255}
\def\Roman#1{\uppercase\expandafter{\romannumeral #1}}
\def\roman#1{\romannumeral #1}
\def\Number#1{\number #1}
\def\unnumberedchapters{\let\makel@bel=\relax \let\chapterlabel=\relax
\let\sectionlabel=\relax \equanumber=-1 }
\def\titlestyle#1{\par\begingroup \interlinepenalty=9999
     \leftskip=0.02\hsize plus 0.23\hsize minus 0.02\hsize
     \rightskip=\leftskip \parfillskip=0pt
     \hyphenpenalty=9000 \exhyphenpenalty=9000
     \tolerance=9999 \pretolerance=9000
     \spaceskip=0.333em \xspaceskip=0.5em
     \iftwelv@\fourteenpoint\else\twelvepoint\fi
   \noindent #1\par\endgroup }
\def\spacecheck#1{\dimen@=\pagegoal\advance\dimen@ by -\pagetotal
   \ifdim\dimen@<#1 \ifdim\dimen@>0pt \vfil\break \fi\fi}
\def\chapter#1{\par \penalty-300 \vskip\chapterskip
   \spacecheck\chapterminspace
   \chapterreset \titlestyle{\chapterlabel \ #1}
   \nobreak\vskip\headskip \penalty 30000
   \wlog{\string\chapter\ \chapterlabel} }

\def\section#1{\par \ifnum\the\lastpenalty=30000\else
   \penalty-200\vskip\sectionskip \spacecheck\sectionminspace\fi
   \wlog{\string\section\ \chapterlabel \the\sectionnumber}
   \global\advance\sectionnumber by 1  \noindent
   {\caps\enspace\chapterlabel \sectionlabel #1}\par
   \nobreak\vskip\headskip \penalty 30000 }
\def\subsection#1{\par
   \ifnum\the\lastpenalty=30000\else \penalty-100\smallskip \fi
   \noindent\undertext{#1}\enspace \vadjust{\penalty5000}}

\def\undertext#1{\vtop{\hbox{#1}\kern 1pt \hrule}}
\def\APPENDIX#1#2{\par\penalty-300\vskip\chapterskip
   \spacecheck\chapterminspace \chapterreset \xdef\chapterlabel{#1}
   \titlestyle{APPENDIX #2} \nobreak\vskip\headskip \penalty 30000
   \wlog{\string\Appendix\ \chapterlabel} }
\def\Appendix#1{\APPENDIX{#1}{#1}}
\def\appendix{\APPENDIX{A}{}}
%
%%%%%%%%%%%%%%%%%%%%%%%%%%%%%%%%%%%%%%%%%%%%%%%%%%%%%%%%%%%%%%%%%%%%%%%%
%
%   Here come macros for equation numbering.
%
\def\eqname#1{\relax \ifnum\the\equanumber<0
     \xdef#1{{\rm(\number-\equanumber)}}\global\advance\equanumber by -1
    \else \global\advance\equanumber by 1
      \xdef#1{{\rm(\chapterlabel \number\equanumber)}} \fi}

\def\eqn#1{\eqno\eqname{#1}#1}

\def\eqinsert#1{\noalign{\dimen@=\prevdepth \nointerlineskip
   \setbox0=\hbox to\displaywidth{\hfil #1}
   \vbox to 0pt{\vss\hbox{$\!\box0\!$}\kern-0.5\baselineskip}
   \prevdepth=\dimen@}}
%

%

%

%
%%%%%%%%%%%%%%%%%%%%%%%%%%%%%%%%%%%%%%%%%%%%%%%%%%%%%%%%%%%%%%%%%%%%%%%%
%   Here come items and lists
%
\def\GENITEM#1;#2{\par \hangafter=0 \hangindent=#1
    \Textindent{$ #2 $}\ignorespaces}
\outer\def\newitem#1=#2;{\gdef#1{\GENITEM #2;}}
\newdimen\itemsize                \itemsize=30pt
\newitem\item=1\itemsize;
\newitem\sitem=1.75\itemsize;     
\newitem\ssitem=2.5\itemsize;     
\outer\def\newlist#1=#2&#3&#4;{\toks0={#2}\toks1={#3}%
   \count255=\escapechar \escapechar=-1
   \alloc@0\list\countdef\insc@unt\listcount     \listcount=0
   \edef#1{\par
      \countdef\listcount=\the\allocationnumber
      \advance\listcount by 1
      \hangafter=0 \hangindent=#4
      \Textindent{\the\toks0{\listcount}\the\toks1}}
   \expandafter\expandafter\expandafter
    \edef\c@t#1{begin}{\par
      \countdef\listcount=\the\allocationnumber \listcount=1
      \hangafter=0 \hangindent=#4
      \Textindent{\the\toks0{\listcount}\the\toks1}}
   \expandafter\expandafter\expandafter
    \edef\c@t#1{con}{\par \hangafter=0 \hangindent=#4 \noindent}
   \escapechar=\count255}
\def\c@t#1#2{\csname\string#1#2\endcsname}
\newlist\point=\Number&.&1.0\itemsize;
\newlist\subpoint=(\alphabetic&)&1.75\itemsize;
\newlist\subsubpoint=(\roman&)&2.5\itemsize;
\newlist\cpoint=\Roman&.&1.0\itemsize;
%

%
%%%%%%%%%%%%%%%%%%%%%%%%%%%%%%%%%%%%%%%%%%%%%%%%%%%%%%%%%%%%%%%%%%%%%%%%
%
%   Here come macros for references, figures & tables.
%
\newcount\referencecount     \referencecount=0
\newif\ifreferenceopen       \newwrite\referencewrite
\newtoks\rw@toks
\def\NPrefmark#1{\attach{\scriptscriptstyle [ #1 ] }}
\let\PRrefmark=\attach
\def\CErefmark#1{\attach{\scriptstyle  #1 ) }}
\def\refmark#1{\relax\ifPhysRev\PRrefmark{#1}\else\NPrefmark{#1}\fi}
\def\crefmark#1{\relax\CErefmark{#1}}
\def\refend{\refmark{\number\referencecount}}
\newcount\lastrefsbegincount \lastrefsbegincount=0
\def\refsend{\refmark{\count255=\referencecount
   \advance\count255 by-\lastrefsbegincount
   \ifcase\count255 \number\referencecount
   \or \number\lastrefsbegincount,\number\referencecount
   \else \number\lastrefsbegincount-\number\referencecount \fi}}
\def\crefsend{\crefmark{\count255=\referencecount
   \advance\count255 by-\lastrefsbegincount
   \ifcase\count255 \number\referencecount
   \or \number\lastrefsbegincount,\number\referencecount
   \else \number\lastrefsbegincount-\number\referencecount \fi}}
\def\refch@ck{\chardef\rw@write=\referencewrite
   \ifreferenceopen \else \referenceopentrue
   \immediate\openout\referencewrite=referenc.texauxil \fi}
%
% In \obeyendofline, we say `\let^^M=\relax
{\catcode`\^^M=\active % these lines must end with %
  \gdef\obeyendofline{\catcode`\^^M\active \let^^M\ }}%
%
% In \ignoreendofline, we say `\let^^M=\relax
{\catcode`\^^M=\active % these lines must end with %
  \gdef\ignoreendofline{\catcode`\^^M=5}}
{\obeyendofline\gdef\rw@start#1{\def\t@st{#1} \ifx\t@st\blankend%
\endgroup \@sf \relax \else \ifx\t@st\bl@nkend \endgroup \@sf \relax%
\else \rw@begin#1
\backtotext
\fi \fi } }
{\obeyendofline\gdef\rw@begin#1
{\def\n@xt{#1}\rw@toks={#1}\relax%
\rw@next}}
\def\blankend{}
{\obeylines\gdef\bl@nkend{
}}
\newif\iffirstrefline  \firstreflinetrue
\def\rwr@teswitch{\ifx\n@xt\blankend \let\n@xt=\rw@begin %
 \else\iffirstrefline \global\firstreflinefalse%
\immediate\write\rw@write{\noexpand\obeyendofline \the\rw@toks}%
\let\n@xt=\rw@begin%
      \else\ifx\n@xt\rw@@d \def\n@xt{\immediate\write\rw@write{%
        \noexpand\ignoreendofline}\endgroup \@sf}%
             \else \immediate\write\rw@write{\the\rw@toks}%
             \let\n@xt=\rw@begin\fi\fi \fi}
\def\rw@next{\rwr@teswitch\n@xt}
\def\rw@@d{\backtotext} \let\rw@end=\relax
\let\backtotext=\relax

\newdimen\refindent     \refindent=30pt
\def\refitem#1{\par \hangafter=0 \hangindent=\refindent \Textindent{#1}}
\def\REFNUM#1{\space@ver{}\refch@ck \firstreflinetrue%
 \global\advance\referencecount by 1 \xdef#1{\the\referencecount}}
\def\refnum#1{\space@ver{}\refch@ck \firstreflinetrue%
 \global\advance\referencecount by 1 \xdef#1{\the\referencecount}\refend}

\def\REF#1{\REFNUM#1%
 \immediate\write\referencewrite{%
 \noexpand\refitem{#1.}}%
\begingroup\obeyendofline\rw@start}
\def\ref{\refnum\?%
 \immediate\write\referencewrite{\noexpand\refitem{\?.}}%
\begingroup\obeyendofline\rw@start}
\def\Ref#1{\refnum#1%
 \immediate\write\referencewrite{\noexpand\refitem{#1.}}%
\begingroup\obeyendofline\rw@start}
\def\REFS#1{\REFNUM#1\global\lastrefsbegincount=\referencecount
\immediate\write\referencewrite{\noexpand\refitem{#1.}}%
\begingroup\obeyendofline\rw@start}
\newcount\figurecount     \figurecount=0
\newif\iffigureopen       \newwrite\figurewrite
\def\figch@ck{\chardef\rw@write=\figurewrite \iffigureopen\else
   \immediate\openout\figurewrite=figures.texauxil
   \figureopentrue\fi}
\def\FIGNUM#1{\space@ver{}\figch@ck \firstreflinetrue%
 \global\advance\figurecount by 1 \xdef#1{\the\figurecount}}
\def\FIG#1{\FIGNUM#1
   \immediate\write\figurewrite{\noexpand\refitem{#1.}}%
   \begingroup\obeyendofline\rw@start}
\def\fig{\FIGNUM\? fig.~\?%
\immediate\write\figurewrite{\noexpand\refitem{\?.}}%
\begingroup\obeyendofline\rw@start}
\def\figure{\FIGNUM\? figure~\?
   \immediate\write\figurewrite{\noexpand\refitem{\?.}}%
   \begingroup\obeyendofline\rw@start}
\def\Fig{\FIGNUM\? Fig.~\?%
\immediate\write\figurewrite{\noexpand\refitem{\?.}}%
\begingroup\obeyendofline\rw@start}
\def\Figure{\FIGNUM\? Figure~\?%
\immediate\write\figurewrite{\noexpand\refitem{\?.}}%
\begingroup\obeyendofline\rw@start}
\newcount\tablecount     \tablecount=0
\newif\iftableopen       \newwrite\tablewrite
\def\tabch@ck{\chardef\rw@write=\tablewrite \iftableopen\else
   \immediate\openout\tablewrite=tables.texauxil
   \tableopentrue\fi}
\def\TABNUM#1{\space@ver{}\tabch@ck \firstreflinetrue%
 \global\advance\tablecount by 1 \xdef#1{\the\tablecount}}
\def\TABLE#1{\TABNUM#1
   \immediate\write\tablewrite{\noexpand\refitem{#1.}}%
   \begingroup\obeyendofline\rw@start}
\def\Table{\TABNUM\? Table~\?%
\immediate\write\tablewrite{\noexpand\refitem{\?.}}%
\begingroup\obeyendofline\rw@start}
%

%
%%%%%%%%%%%%%%%%%%%%%%%%%%%%%%%%%%%%%%%%%%%%%%%%%%%%%%%%%%%%%%%%%%%%%%%%
%
%   Here come macros for memos & letters.
%
\def\masterreset{\global\pagenumber=1 \global\chapternumber=0
   \ifnum\the\equanumber<0\else \global\equanumber=0\fi
   \global\sectionnumber=0
   \global\referencecount=0 \global\figurecount=0 \global\tablecount=0 }
\def\FRONTPAGE{\ifvoid255\else\vfill\penalty-2000\fi
      \masterreset\global\frontpagetrue
      \global\lastf@@t=0 \global\footsymbolcount=0}

\def\paperstyle{\letterstylefalse\normalspace\papersize}
\def\letterstyle{\letterstyletrue\singlespace\lettersize}
\def\papersize{\hsize=35pc\vsize=48pc\hoffset=1pc\voffset=6pc
               \skip\footins=\bigskipamount}
\def\lettersize{\hsize=6.5in\vsize=8.5in\hoffset=0in\voffset=1in
   \skip\footins=\smallskipamount \multiply\skip\footins by 3 }
\paperstyle   %  This is the default
%
% % % % % % % % % % % % % % % % % % % % % % % % % % % % % % % % % % % %
%
\def\MEMO{\letterstyle\FRONTPAGE \letterfrontheadline={\hfil}
    \line{\quad\fourteenrm FNAL MEMORANDUM\hfil\twelverm\the\date\quad}
    \medskip \memod@f}

\def\memit@m#1{\smallskip \hangafter=0 \hangindent=1in
      \Textindent{\caps #1}}
\def\memod@f{\xdef\to{\memit@m{To:}}\xdef\from{\memit@m{From:}}%
     \xdef\topic{\memit@m{Topic:}}\xdef\subject{\memit@m{Subject:}}%
     \xdef\rule{\bigskip\hrule height 1pt\bigskip}}
\memod@f
\newskip\lettertopfil
\lettertopfil = 0pt plus 1.5in minus 0pt
\newskip\letterbottomfil
\letterbottomfil = 0pt plus 2.3in minus 0pt
\newskip\spskip \setbox0\hbox{\ } \spskip=-1\wd0
\def\addressee#1{\medskip\rightline{\the\date\hskip 30pt} \bigskip
   \vskip\lettertopfil
   \ialign to\hsize{\strut ##\hfil\tabskip 0pt plus \hsize \cr #1\crcr}
   \medskip\noindent\hskip\spskip}
\newskip\signatureskip       \signatureskip=40pt
\def\signed#1{\par \penalty 9000 \bigskip \dt@pfalse
  \everycr={\noalign{\ifdt@p\vskip\signatureskip\global\dt@pfalse\fi}}
  \setbox0=\vbox{\singlespace \halign{\tabskip 0pt \strut ##\hfil\cr
   \noalign{\global\dt@ptrue}#1\crcr}}
  \line{\hskip 0.5\hsize minus 0.5\hsize \box0\hfil} \medskip }

\def\endletter{\ifnum\pagenumber=1 \vskip\letterbottomfil\supereject
\else \vfil\supereject \fi}
\newbox\letterb@x
\def\lettertext{\par\unvcopy\letterb@x\par}
\def\multiletter{\setbox\letterb@x=\vbox\bgroup
      \everypar{\vrule height 1\baselineskip depth 0pt width 0pt }
      \singlespace \topskip=\baselineskip }
\def\letterend{\par\egroup}
%
%%%%%%%%%%%%%%%%%%%%%%%%%%%%%%%%%%%%%%%%%%%%%%%%%%%%%%%%%%%%%%%%%%%%%%%
%
%   Here come macros for title pages.
%
\newskip\frontpageskip
\newtoks\pubtype
\newtoks\Pubnum
\newtoks\pubnum
\newif\ifp@bblock  \p@bblocktrue
\def\PH@SR@V{\doubl@true \baselineskip=24.1pt plus 0.2pt minus 0.1pt
             \parskip= 3pt plus 2pt minus 1pt }
\def\PHYSREV{\paperstyle\PhysRevtrue\PH@SR@V}
\def\titlepage{\FRONTPAGE\paperstyle\ifPhysRev\PH@SR@V\fi
   \ifp@bblock\p@bblock\fi}
\def\nopubblock{\p@bblockfalse}
\def\endpage{\vfil\break}
\frontpageskip=1\medskipamount plus .5fil
\pubtype={\tensl Preliminary Version}
%\Pubnum={$\caps FERMILAB - Pub - \the\pubnum $}
%\Pubnum={$\rm FERMILAB-Pub-\the\pubnum $}
\pubnum={0000}
\def\p@bblock{\begingroup \tabskip=\hsize minus \hsize
   \baselineskip=1.5\ht\strutbox \topspace-2\baselineskip
   \halign to\hsize{\strut ##\hfil\tabskip=0pt\crcr
%   \the\Pubnum\cr \the\date\cr \the\pubtype\cr}\endgroup}
   \the\Pubnum\cr \the\date\cr}\endgroup}
%   \the\date\cr}\endgroup}

%
\def\title#1{\vskip\frontpageskip \titlestyle{#1} \vskip\headskip }
\def\author#1{\vskip\frontpageskip\titlestyle{\twelvecp #1}\nobreak}

\def\address#1{\par\kern 5pt\titlestyle{\twelvepoint\it #1}}
\def\andaddress{\par\kern 5pt \centerline{\sl and} \address}

\def\abstract{\vskip\frontpageskip\centerline{\fourteenrm ABSTRACT}
              \vskip\headskip }

%
%
%%%%%%%%%%%%%%%%%%%%%%%%%%%%%%%%%%%%%%%%%%%%%%%%%%%%%%%%%%%%%%%%%%%%%%%%
%   Miscellaneous macros
%

\def\\{\relax\ifmmode\backslash\else$\backslash$\fi}
\def\globaleqnumbers{\relax\ifnum\the\equanumber<0%
\else\global\equanumber=-1\fi}

\def\journal#1&#2(#3){\unskip, \sl #1~\bf #2 \rm (19#3) }

\def\topspace{\hrule height 0pt depth 0pt \vskip}

\let\int=\intop         
\def\prop{\mathrel{{\mathchoice{\pr@p\scriptstyle}{\pr@p\scriptstyle}{
                \pr@p\scriptscriptstyle}{\pr@p\scriptscriptstyle} }}}
\def\pr@p#1{\setbox0=\hbox{$\cal #1 \char'103$}
   \hbox{$\cal #1 \char'117$\kern-.4\wd0\box0}}
\def\lsim{\mathrel{\mathpalette\@versim<}}
\def\gsim{\mathrel{\mathpalette\@versim>}}
\def\@versim#1#2{\lower0.2ex\vbox{\baselineskip\z@skip\lineskip\z@skip
  \lineskiplimit\z@\ialign{$\m@th#1\hfil##\hfil$\crcr#2\crcr\sim\crcr}}}
\def\leftrightarrowfill{$\m@th \mathord- \mkern-6mu
	\cleaders\hbox{$\mkern-2mu \mathord- \mkern-2mu$}\hfil
	\mkern-6mu \mathord\leftrightarrow$}
\def\lrover#1{\vbox{\ialign{##\crcr
	\leftrightarrowfill\crcr\noalign{\kern-1pt\nointerlineskip}
	$\hfil\displaystyle{#1}\hfil$\crcr}}}
%
% % % % % % % % % % % % % % % % % % % % % % % % % % % % % % % % % % % %
%
%   Finally, some bug fixings.
%
\let\sec@nt=\sec
\def\sec{\relax\ifmmode\let\n@xt=\sec@nt\else\let\n@xt\section\fi\n@xt}
\def\obsolete#1{\message{Macro \string #1 is obsolete.}}
\def\firstsec#1{\obsolete\firstsec \section{#1}}
\def\firstsubsec#1{\obsolete\firstsubsec \subsection{#1}}
\def\thispage#1{\obsolete\thispage \global\pagenumber=#1\frontpagefalse}
\def\thischapter#1{\obsolete\thischapter \global\chapternumber=#1}
\def\nextequation#1{\obsolete\nextequation \global\equanumber=#1
   \ifnum\the\equanumber>0 \global\advance\equanumber by 1 \fi}
\def\BOXITEM{\afterassigment\B@XITEM\setbox0=}
\def\B@XITEM{\par\hangindent\wd0 \noindent\box0 }
%

%%%%%%%%%%%%%%%%%%%%%%%%%%%%%%%%%%%%%%%%%%%%%%%%%%%%%%%%%%%%%%%%%%%%%%%%
%   That's about it
%
\catcode`@=12 % at signs are no longer letters
\message{ by V.K.}
\everyjob{\input myphyx }

\hsize=5.5truein
\voffset=-.3truein
\nopagenumbers
\baselineskip=14pt
\overfullrule=0pt
\font\Bigbf=cmbx12 scaled 1440
\voffset= -.3truein
\nopagenumbers
\line{\hfil WM-93-108}
\line{\hfil July 1993}
\vskip .75truein
\centerline{\Bigbf Precise Vacuum Stability Bound}
\smallskip
\centerline{\Bigbf in the Standard Model}
\vskip 1.0truein
\centerline{\bf Marc Sher}
\vskip .5truein
\centerline{Physics Dept}
\centerline{College of William and Mary}
\centerline{Williamsburg VA 92093}
\bigskip\bigskip
\centerline{\bf Abstract}
\smallskip
In the standard model, a lower bound to the Higgs mass (for a given top quark
mass) exists if one requires that the standard model vacuum be stable.  This
bound is calculated as precisely as possible, including the most recent values
of the strong and electroweak couplings, corrected two-loop beta functions and
radiative corrections to the Higgs and top quark masses.  In addition to
being somewhat more precise, this work differs from previous calculations in
that the bounds are given in terms of the poles of the Higgs and top quark
propagators, rather than, for example, the ``$\overline{MS}$ top quark mass''.
This difference can be as large as $6-10$ GeV for the top quark mass,
which corresponds to as much as $15$ GeV for the lower bound to
the Higgs mass.  I
concentrate on top quark masses between $130$ and $150$ GeV, and for
$\alpha_s(M_Z)=0.117$ find that (over that range) $m_H\ >\
75\ {\rm GeV}+ 1.64(m_t-140\ {\rm GeV})$.  This result increases (decreases)
by $3$ GeV if the strong coupling decreases (increases) by $0.007$, and is
accurate to $1$ GeV in $m_t$ and $2$ GeV in $m_H$.
If one allows for the standard model vacuum to be unstable, then
weaker bounds can be obtained--these are also discussed.
\vfill
\eject
\pagenumbers

\REF\lep{See W.J. Marciano, {\it The Fermilab Meeting},  Annual Meeting of the
Division of Particles and Fields of the APS, Batavia, IL, 1992, edited by C.H.
Albright, et al. (World Scientific, Singapore, 1992) p.185.}
 \REF\rep{M. Sher,
Phys. Reports {\bf 179} (1989) 273.} \REF\lsz{M. Lindner, M. Sher and H.W.
Zaglauer, Phys. Lett. {\bf 228B} (1989) 139.} \REF\av{P. Arnold
and Vokos, Phys. Rev. {\bf D44} (1991) 3620; G.W. Anderson, Phys. Lett. {\bf
243B} (1990) 265.}
\REF\els{J. Ellis, A. Linde and M. Sher, Phys. Lett. {\bf 252B} (1990) 203.}
\REF\fjse{C. Ford, D.R.T. Jones, P.W. Stephenson and M.B. Einhorn, Nucl. Phys.
{\bf B395} (1993) 62.}
\REF\uf{H. Aronson, D.J. Castano, B. Kesthelyi, S. Mikaelian, E.J. Piard,
P. Ramond and B.D. Wright, Phys. Rev. {\bf D46} (1992) 3945.}
\REF\sz{ A. Sirlin and R. Zucchini, Nucl. Phys. {\bf B226} (1986) 389.}
\REF\supp{M. Maggiore and M. Shifman, Nucl. Phys. {\bf B380}
(1992) 22; V.I. Zakharov, Nucl. Phys. {\bf B383} (1992) 218.}
\REF\nosupp{J.M. Cornwall, Phys. Lett. {\bf B243} (1990) 271;
H. Goldberg, Phys. Lett {\bf B246} (1990) 445, Phys. Rev. {\bf D45}
 (1992) 2945; E.N. Argyres, R.H.P. Kleiss and C.G. Papadopoulos, \
Nucl. Phys. {\bf B391} (1993) 42, {\bf B391} (1993) 57.}

The highly successful standard model still has two missing ingredients: the top
quark and the Higgs boson.  The Higgs boson mass can have any value up to
approximately $600-800$ GeV, whereas the top quark mass is restricted by
analysis
of radiative corrections at LEP [\lep] to values below approximately $200$ GeV.
The strongest restriction on the allowed mass values comes from the requirement
of vacuum stability.  If the top quark is too heavy, then its
contribution (which is negative) to the beta function for the
scalar self-coupling is sufficiently large to drive that coupling
to negative values, destabilizing the vacuum.
Requiring that the
standard model vacuum be the ground state of the model leads to a lower bound
on the Higgs mass (for a given top quark mass); for top quark masses between
$130$ and $150$ GeV, this Higgs mass lower bound ranges from $65$ to $100$ GeV.
Even if one allows for the standard model vacuum to be unstable, weaker bounds
can be obtained from the requirement that the universe arrive in the standard
model vacuum and remain there for ten billion years.  One must also require
that
high energy cosmic ray collisions not touch off the decay of the vacuum.

These bounds are all discussed in detail in Ref. [\rep].  After that article
appeared, several improvements in the bounds were made.  Lindner et al.[\lsz]
calculated the vacuum stability bound using two-loop beta function, anomalous
dimensions and Higgs mass correction terms.  Arnold and Vokos[\av] showed
that the requirement that thermal fluctuations in the early
Universe not be strong enough to force the Universe out of
the correct vacuum is stronger than
the requirement that it stay there for at least ten billion years.  Ellis et
al.[\els] considered the question of cosmic ray nucleation of the transition,
including recent results on multiparticle production of Higgs bosons at high
energies.

More recently, Ford, Jones, Stephenson and Einhorn[\fjse] performed a detailed
study of the effective potential and the renormalization group through
two-loops.  They repeated the vacuum stability calculation of Lindner et al.
In addition to putting previous calculations on a sounder theoretical
footing, their work was an improvement in several respects.  They corrected a
typographical error in the two-loop beta function for the scalar self-coupling
which appeared in the original calculation of this quantity, and they used much
more precise values for the gauge coupling constants.  However, in their
calculation, they used tree-level values for the top quark and Higgs masses
and, as we will see, corrections to the Higgs mass can be 1-2 GeV and
corrections to the top quark mass can be as large as 8 GeV.

In this Letter, I will recalculate all of the above bounds, including the
typographical error correction, precise values of the gauge coupling
constants and higher order corrections to the  Higgs and top quark masses.
In particular, results will be given for the allowed region
of masses, where masses are give by the
pole of the Higgs propagator and the pole of the top quark propagator (all
previous calculations just determined the $\overline{MS}$ value of the top
quark mass at some scale).  These results will be accurate (given the strong
coupling constant) to within a GeV.

Why should one consider determining the bounds to such precision?  Recently,
several top quark candidates have been seen at CDF and D0.  Although not enough
statistics exists for a claim of a top quark discovery, these events combined
with the results from electroweak radiative corrections do point towards a top
quark mass between $130$ and $150$ GeV.  Over this region, the stability bound
corresponds to a lower bound on the Higgs mass which varies from approximately
$65$ GeV (for a top mass of $130$ GeV) to $100$ GeV (for a top mass of $150$
GeV).  This bound is precisely in the range to be explored by LEP II within the
next two to three years, and one would like to have a precise determination.
I will thus focus on the top quark mass range from $130$ to $150$ GeV.

The bound arising from vacuum stability is discussed in detail in Ref. \rep.
In short, it arises because the top quark Yukawa coupling gives a negative
contribution to the beta function of the scalar self-coupling.  If the top
quark is sufficiently heavy, this causes the scalar self-coupling to decrease
with scale, eventually becoming negative.  At this point, the value of
the effective potential becomes negative and drops lower than the
standard model vacuum.   The calculation
can be simplified greatly by noting that, {\it for the top quark mass range of
interest}, the instability only sets in at scales much larger than $10$ TeV
(see, for example, the figure in Ref. \lsz).  At these scales, the quadratic
term
in the potential is utterly negligible, and the potential can be written ({\it
at large scales}) as $$V={1\over 4}\lambda(t) G^4(t) \phi^4\eqn\tree$$ where
$$\eqalign{G&\equiv \exp\left(-\int_0^t dt'\ {\gamma\over 1-\gamma}\right)\cr
{d\lambda\over dt}&={\beta\over 1-\gamma}\cr
{dg_k\over dt}&={\beta_{g_k}\over 1-\gamma}\cr}\eqn\gmm$$ Here, $g_k$
refers to the gauge and Yukawa couplings, $t\equiv \log(\phi/M)$, and the
 $\beta$ and $\gamma$ functions depend on all of the couplings in the model.
Clearly, the standard model vacuum will be unstable if $\lambda(t)$ becomes
negative.

To determine if the standard model vacuum is unstable, one simply has to start
with initial values of $\lambda, g, g^{\prime}, g_s$ and $g_Y$ at some scale
which I will choose to be $M_Z$.  The equations
$$\eqalign{{d\lambda\over dt}&=\beta\cr {d{g_k}\over
dt}&=\beta_{g_k}\cr}\eqn\equations$$
are then integrated up to a large scale (the instability will typically set in
at some point between $10^6$ and $10^{10}$ GeV) to see if $\lambda$ becomes
negative.  For top quark masses in the range of interest, the value of $\beta$
at low scales is always negative.  Since the top quark Yukawa coupling itself
falls with scale, the value of $\beta$ at high scales is typically positive.
As a result, $\lambda(t)$ will fall with scale until some minimum is reached,
and then rise.  If this minimum is above zero, the standard model vacuum is
stable (see Ref. \fjse\ for a very clear discussion).  Note that the anomalous
dimension factors have been absorbed into the definition of $\phi$, since the
precise location of the instability is irrelevant[\fjse].  Two-loop beta
functions are used throughout; they are explicitly written out in Ref. [\fjse].

Thus, one can determine whether the vacuum is unstable for initial values of
$\lambda, g, g^{\prime}, g_s$ and $g_Y$ at the scale $M_Z$.  (The
$\overline{MS}$ renormalization scheme is used throughout.)  Now, these
parameters must be related to measureable quantities.  The gauge couplings
in the ${\overline{MS}}$ scheme are given by[\uf]
$$\eqalign{g(M_Z)&=.6502\ \pm\ .002\cr
g^{\prime}(M_Z)&=.3578\ \pm\ .001\cr}
\eqn\gaugecouplings$$
For the strong coupling constant, we will choose a range of values given by
$\alpha_s=.117\pm .007$.

 The two remaining
parameters are $\lambda$ and $g_Y$.  The relationship between $\lambda$ in the
$\overline{MS}$ scheme and the Higgs mass was calculated in Ref. [\sz].  It is
given by $$\lambda(M_Z)={G_{\mu}\over\sqrt{2}}M_H^2[1+\delta(M_Z)]\eqn\deldef$$
where $\delta(M_Z)$ contains the radiative corrections and is given explicitly
in Ref. [\uf].  However, $M^2_H$ is {\it not}, strictly speaking, the Higgs
mass.
{}From Ref. [\sz], one can see that it is the curvature of the potential at the
minimum.  The inverse scalar propagator can be written as
$-iG^{-1}(p)=p^2-m^2-\Sigma(p)$, where $\Sigma(p)$ is the scalar self-energy.
Expanding this about $p^2=m^2$, and using the fact that the curvature of the
potential is the inverse propagator at {\it zero} external momentum, one easily
finds that the pole of the scalar propagator occurs at
$$m^2_H=\left(1+{d\Sigma\over
dp^2}\big|_{p^2=m^2}\right)M^2_H\eqn\higgsmass$$
The $p$-dependent part of $\Sigma$ can easily be found, and this correction
factor (which is generally less than one percent) determined.

Finally, we need to relate the top quark mass to $g_Y(M_Z)$.  In Ref. [\fjse],
the expression $m_t = g_Y(M_Z)\sigma/\sqrt{2}$ was used, where $\sigma$ is the
vacuum expectation value of the Higgs field (note that $\sigma$ receives
extremely small radiative corrections, as shown in [\uf]). In Ref. [\lsz],  it
was argued that an appropriate definition of the top quark mass was half the
energy needed to pair produce them, i.e. $m_t= g_Y(2M_t)\sigma/\sqrt{2}$.  Each
of these definitions, however, neglects a very large QCD correction.  The most
appropriate definition of the top quark mass is the pole of the propagator,
which is related to the $\overline{MS}$ definition of the mass by the
expression (see Ref. [\uf] for an extensive discussion): $${M_t\over
m_t(M_t)}=1+{4\over 3}{\alpha_s(M_t)\over \pi}
+10.91\left({\alpha_s(M_t)\over\pi}\right)^2\eqn\topmass$$
This correction can be very large, as much as $6$ percent, or $8-10$ GeV over
the range of interest.  The correction swamps most of the other correction
factors.  Due to uncalculated three-loop effects, it is itself uncertain by
roughly one GeV.  At present, one therefore cannot calculate the bound any
more precisely.

Putting all of this together, I find the result in Fig. [1].  The stability
bound has been listed for three values of the strong coupling constant, which
span the likely range of values.  Changing the SU(2) and U(1) couplings over
the allowed range of their values changes the Higgs mass bound by only $20$
MeV.  Although the curves are not precisely linear, a linear function can be
found which is valid to the accuracy stated in the previous paragraph ($1$
GeV in the top quark mass, or $2$ GeV in the Higgs mass) {\it over the range
from $130$ to $150$ GeV}: $$ m_H\ >\ 75\ + \ 1.64 (m_t-140)\ -\
3\left({\alpha_s-.117\over .007}\right),\eqn\result$$ in units of GeV.  For
the ranges of top mass between $120$ and $130$ GeV, or betwenn $150$ and
Note that this differs by approximately $10$ GeV from the recent
work of Ref. [\fjse] due to the more
$160$ GeV, this formula underestimates the bound by approximately $2$ GeV.
precise definition of the top quark mass.

I now turn to the other bounds discussed in the introduction.  If the
vacuum is unstable, then its lifetime must be less than approximately ten
billion years.  The techniques for calculating the lifetime of a metastable
vacuum state are reviewed in detail in Ref. [\rep].
In natural units, the volume of our past light cone is $M^{-4}_Ze^{404}$, and
the
nucleation rate per unit time per unit volume is $\phi^4_oe^{-S_E}$, where
$\phi_o$ is the value of the field just after the transition and $S_E$ is the
Euclidean action.  Although these quantities can be found numerically,
 an extremely accurate
analytical approximation can be found
\REF\arn{P. Arnold, Phys. Rev. {\bf D40} (1989) 613.}
in the work of Arnold[\arn], who shows that the requirement that the lifetime
exceed ten billion years is essentially the same as the requirement that
$$\max_{\lambda(t)<0}\left(\phi^4e^{-S}\right) \lsim M^4_Ze^{-404}
\eqn\arnoldbound$$
Here, $S=8\pi^2/3|\lambda(t)|$, and  one  finds the maximum value of the term
in
parentheses over the range of $t$ for which $\lambda$ is negative.  If this
range extends past the unification scale, it will be cut off at that point;
this
has very little effect on the results.  This requirement can be written as
$$\min_{\lambda(t)<0}\left({8\pi^2\over 3|\lambda(t)|}-4t\right)\ >\ 404.$$
A stronger requirement can be obtained by considering
the possibility that  thermal fluctuations in the early
universe cause the universe to fall into the true vacuum.    Arnold and Vokos,
following work of Anderson, showed that this
condition is stronger than the above, and they find that one must have
$$\min_{\lambda(t)<0}\left(1+{6\pi\xi(t)\over\lambda(t)}\right) \lsim 232,$$
where $12\xi(t)\equiv 9g^2(t)/4+3g^{\prime 2}/4+3g^2_Y(t)+6\lambda(t)$. These
bounds are plotted in Fig. [2].

Finally, if the vacuum is unstable, yet satisfies the bounds in the previous
paragraph, one still must consider the possibility that high energy cosmic ray
collisions touch off the transition (as they do, for example, in bubble
chambers).  As first pointed out by Arnold[\arn], it is not simply sufficient
for
the cosmic ray collision to have enough center of mass energy to get over the
potential energy barrier.  Rather, the collision must also produce a coherent
state of a large number of Higgs bosons.  Here, the controversy concerning the
cross section for multiparticle production at high energies becomes relevant.
Some have argued[\supp] that the cross section is suppressed by a factor of
$\exp(-8\pi^2/g^2)$; some have argued[\nosupp] that the cross section is at or
near the unitarity limit.  If it turns out that the cross section is
exponentially suppressed, then cosmic rays will not touch off the transition
under any circumstances.  However, if the cross section is near the unitarity
limit, then they might induce the transition if they have enough energy to
produce a very large number of Higgs bosons.  This is discussed in detail in
Ref. [\els].  Using the formalism described there, and assuming that the cosmic
ray collision creates a spherical distribution of Higgs bosons, I find that
only
a very small region of parameter space allowed by the bounds in the above
paragraph can be excluded\foot{If the collision produces an asymmetric
collision, then  a larger region of parameter space will be excluded, but there
are greater uncertainties.}

In this Letter, I have calculated the vacuum stability bound in the
standard model as precisely as possible.  Higher precision would require
calculating the three-loop QCD corrections to the top quark mass;
and since the uncertainty in the strong coupling constant will dominate
such corrections, such a calculation would not be productive.  One can
see from Fig. 1 that, if the top quark mass is between $130$ and $150$
GeV, this bound will be definitively tested at LEP II.  Should the bound
be violated, one could consider the possibility that our vacuum is
unstable; Fig. 2 shows how the bounds are  weaker if this
possibility is allowed.  Failure to satisfy these latter bounds will rule out
the minimal standard model; extensions of the standard model will have
significantly different bounds[\rep] \endpage

\par \penalty-400 \vskip\chapterskip
   \spacecheck\referenceminspace \immediate\closeout\referencewrite
   \referenceopenfalse
   \line{\fourteenrm\hfil REFERENCES\hfil}\vskip\headskip
   \input referenc.texauxil

\FIG\fone{Vacuum stability bound in the standard model for three values of the
strong coupling constant $\alpha_s$ (in the $\overline{MS}$ scheme).  In the
region below the lines, the standard model vacuum is unstable.}

\FIG\ftwo{For $\alpha_s=0.117$, the vacuum stability bound from Fig. 1 is
plotted
as the solid line[S].  Below this line, the standard model vacuum is unstable.
Below the dashed line[TF], thermal fluctuations in the early universe will
cause the
universe to enter the true vacuum; below the dot-dashed line[QF], the lifetime
of
our vacuum is less than ten billion years.  Thus, the region below the dashed
line is ruled out.  In the region below the dotted line[CR], cosmic rays would
have
induced the transition {\it if} the multiparticle production rate at high
energies is at or near the unitarity limit; the region below the dotted line
thus may be ruled out as well.}\endpage\par \penalty-400 \vskip\chapterskip
   \spacecheck\referenceminspace \immediate\closeout\figurewrite
   \figureopenfalse
   \line{\fourteenrm\hfil FIGURE CAPTIONS\hfil}\vskip\headskip
   \input figures.texauxil
    \end